\documentclass[aip,pof,reprint]{revtex4-1}
\usepackage[ansinew]{inputenc} 
\usepackage[T1]{fontenc} 
\usepackage[english]{babel} 
\usepackage{graphicx}
\usepackage{xcolor}
\usepackage{bm} 
\usepackage{amsmath} 
\usepackage{amsfonts}
\usepackage{amssymb}
\usepackage[
            pdfpagemode = UseNone,
pdfborder={0 0 0}, 
colorlinks=true, 
linkcolor=blue,
citecolor=blue,
urlcolor=blue
]{hyperref} 

\begin{document}

\title{Leapfrogging Kelvin waves}
\author{N. Hietala}
\email{niklas.hietala@aalto.fi}
\author{R. Hänninen}
\affiliation{Low Temperature Laboratory, Department of Applied Physics, Aalto University, PO Box 15100, FI-00076 AALTO, Finland}
\author{H. Salman}
\affiliation{School of Mathematics, University of East Anglia, Norwich Research Park, Norwich NR4 7TJ, United Kingdom}
\author{C.~F. Barenghi}
\affiliation{Joint Quantum Centre Durham-Newcastle, School of Mathematics and Statistics, Newcastle University, Newcastle upon Tyne NE1 7RU, United Kingdom}
\date{\today}

\begin{abstract}
Two vortex rings can form a localized configuration whereby they continually pass through one another in an alternating fashion. This phenomenon is called leapfrogging. Using parameters suitable for superfluid helium-4, we describe a recurrence phenomenon that is similar to leapfrogging, which occurs for two coaxial straight vortex filaments with the same Kelvin wave mode. For small-amplitude Kelvin waves we demonstrate that our full Biot-Savart simulations closely follow predictions obtained from a simplified model that provides an analytical approximation developed for nearly parallel vortices. Our results are also relevant to thin-cored helical vortices in classical fluids.
\end{abstract}
\maketitle

\section{Introduction}
The mathematical foundation of vortex dynamics was laid down by Helmholtz\cite{hel58, mel10}, who subsequently applied his theory to study the propagation of vortex rings. In his work, he suggested that two vortex rings moving along the same axis would thread each other in an alternating fashion. The study of vortex rings was also taken up by Lord Kelvin who  contributed significantly to our understanding of the motion of vortices in general.
Following the works of Helmholtz and Kelvin, the leapfrogging motion of vortex rings has been studied in more detail for classical fluids \cite{dys93, hic22, bor13} and also for superfluids \cite{wac14, cap14}.

Leapfrogging of vortex rings is an interesting example of a recurrence phenomenon involving two vortices. In this work, we will give another example of leapfrogging, which can resemble the motion of two coaxial vortex rings. Part of our motivation is to understand the interaction of Kelvin waves on quantized superfluid vortices. It has been argued that Kelvin waves, which are helical perturbations of a straight vortex, are 
important for the energy dissipation in superfluid turbulence at very low temperatures \cite{vin01,koz09,bou11}. However, a large body of analytical results and a number of numerical studies rely on the simplifying assumption of neglecting the interaction of the Kelvin waves between different vortex filaments and focus on how Kelvin waves evolve on a single filament. Despite this, the justification of these assumptions on which many of these theories are based is not fully established.

In this work we will uncover a novel type of interaction between adjacent vortex filaments that can play an important role in our understanding of how energy is transferred across different length scales. We note that some Kelvin wave phenomena bear resemblance to the motion of vortex rings. For example, a superfluid vortex ring experiencing a counterflow (the relative velocity of normal fluid and superfluid \cite{Donnelly}) through it will either shrink or grow depending on the direction of the counterflow. Similarly, the amplitude of a Kelvin wave will either decrease or increase depending on the amount of the counterflow along the vortex axis \cite{bar04}. 
On the other hand, a vortex with a large-amplitude Kelvin wave has a shape that corresponds to a tightly wound helix that, in some approximate sense, bears resemblance to a stacked row of vortex rings. Although the correspondence is not exact, in this work we propose that two stacked rows of vortex rings mimics the motion of two vortices with large-amplitude Kelvin waves. It turns out that this analogy provides a qualitative understanding of the observed dynamics. Interestingly, we will show that a form of vortex leapfrogging persists even when the Kelvin wave amplitudes are small.

Aside from their importance for superfluid turbulence, helical vortices are also important for classical fluid dynamics. For example, the wake behind rotors can be treated as one or many interlaced helical vortices. Circumstances where helical vortices are relevant include wakes behind propellers, wind turbines, or helicopter blades \cite{jai00, sta05, oku07, iva10, fel11, sar14, lew14, del15, nem15, sel15}. Moreover, experiments have shown how the adjacent turns of two helices may contract and expand in a manner that resembles leapfrogging of vortex rings \cite{sta05, fel11}. 

In this work, we study the leapfrogging of helical vortices, where the contraction and expansion occurs along the entire helix as opposed to individual turns. In ordinary viscous fluids, it is expected that the interaction between helical vortices will eventually lead to the merging of the vortices \cite{del15, sel15}. However, in our superfluid context, viscous effects and the dynamics of the vortex core are irrelevant (the core has a negligible atomic scale). In this scenario, the problem of the interaction of the Kelvin waves acquires its simplest possible form.

\section{Methods \label{Sec_Methods}}
We model the dynamics of helical superfluid vortices using a vortex filament model in which the vortices are described as discretized space curves ${\bm s}(\xi,t)$ that are parameterised by their arc length $\xi$. 
We use parameters typical for helium-4 experiments:
Each vortex has circulation  $\kappa = 0.0997$ mm$^2$/s and a core diameter $a_0 \approx 10^{-7}$ mm \cite{Donnelly}. Since the core diameter is several orders of magnitude smaller than any other length scale in the system, for example the characteristic intervortex separation in the experiments, this justifies the use of the filament model. 
With this model, a filament is discretized by a finite set of points along the curve ${\bm s}(\xi,t)$. Each vortex point then evolves according to the Biot-Savart law
\begin{equation}\label{e.bs}
\dot{\bm s} =
\frac{\kappa}{4\pi} \ln\left(\frac{2\sqrt{l_{+}l_{-}}}{e^{1/2}a_0}\right) \hat{\bm s}'\times {\bm s}'' + 
\frac{\kappa}{4\pi}\int' \frac{({\bm s}_1-{\bm s})\times {\rm d}{\bm s}_1}
{\vert {\bm s}_1-{\bm s}\vert^3}\, \text{,} 
\end{equation}
where ${\bm s}_1 = {\bm s}(\xi_1,t)$, the overdot denotes a time derivative, and a prime to the variable ${\bf s}$ denotes differentiation with respect to arc length. 
The first term arises from a commonly adopted regularisation of the singular integrand in the expression for the velocity in the Biot-Savart integral in which the contribution from a small interval $[ {\bm s}(\xi-l_-),{\bm s}(\xi+l_+) ]$ is isolated around the point at which  the velocity is calculated \cite{sch85}.
\begin{figure}[bt]
\includegraphics[width=0.9\textwidth]{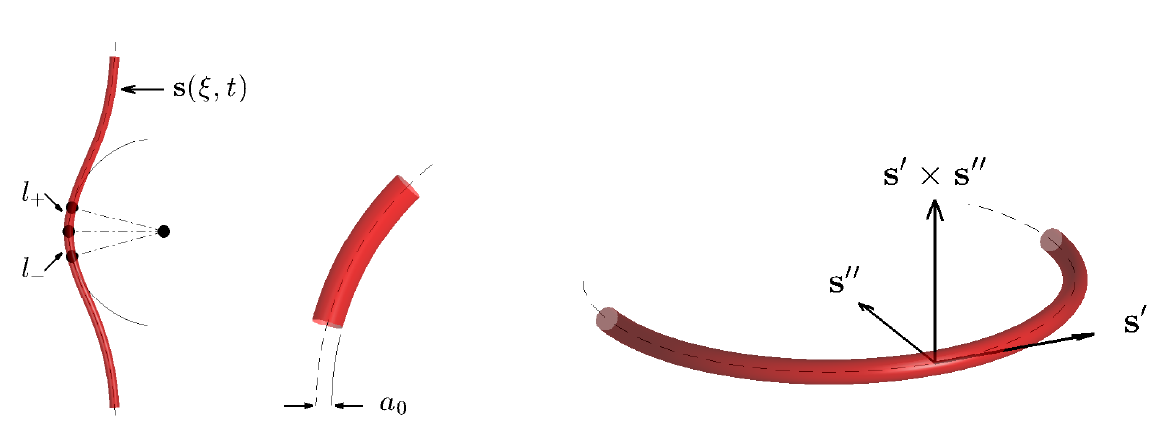}
\caption{
Key length scales used in the vortex filament model. Also shown is the right-handed local orthogonal coordinate system prescribed by the local vectors ${\bf s}'$, ${\bf s}''$ and ${\bf s}' \times {\bf s}''$.}

\label{fig:schematic}
\end{figure}
The Biot-Savart integral contained in the second term contains the contribution from the remaining segment of the filament plus the contributions from all the other vortices. Henceforth, we denote an integral that excludes a segment containing the divergent contribution with a prime. The above approximation is valid when $a_0 \ll l_\pm \ll 1/c$, where $c$ is the local curvature. 
In our simulations, $l_\pm$ were chosen to correspond to the distances to the two nearby points on the discretized curve, respectively. Figure \ref{fig:schematic} illustrates the parametrization and the key length scales used in the model. 

In this work we focus on the motion of vortices at zero temperature in the absence of any normal fluid, which allows us to neglect mutual friction that would otherwise act on the vortices \cite{Donnelly}; in practice, this regime refers to helium-4 experiments at temperatures below $1~{\rm K}$. Above this temperature, mutual friction acts to damp Kelvin waves along vortex filaments. Without the normal component, the dynamics of the vortices are equivalent to the motion of vortices in a classical ideal fluid but with the constraint that the circulation is quantized in units of $h/m_4$, where $m_4$ is the atomic mass of $^4$He.

A Kelvin wave is a helical perturbation of a straight vortex. We choose our coordinate system such that the principal axis of the filament is aligned 
along the $z$~axis. Since this work deals with vortex filaments with helical shape, the $x$ and $y$ coordinates are single-valued functions of the $z$ coordinate and we can therefore introduce a complex function
\begin{equation} \label{eq.w}
w(z) = x(z) + iy(z) \text{.}
\end{equation} 
Then the equation of a helical vortex is given by $w(z) = A \exp \bm( i(kz + \phi)\bm)$. Here $A$ is the amplitude of the wave, $k$ is the wave number, and $\phi$ is a phase that determines the orientation of the helix. For an infinitely long straight vortex, $k$ is any real number. In this work we will simulate vortices in a domain with periodic boundary conditions along the $z$ coordinate direction. In this case, allowable wave numbers correspond to $k = 2 \pi m / L_z$, where $m$ is an integer and $L_z$ is the period along $z$.

In general, a curve that is a single-valued function of $z$ can be expressed as a linear combination of Kelvin waves given by
\begin{equation}
w(z) = \sum_k \alpha_k \exp \bm( i(kz + \phi_k) \bm) \text{,}
\end{equation}
where $\alpha_k$ and $\phi_k$ are the amplitude and phase of the $k$th mode, respectively.
This allows us to decompose a general perturbation in terms of its Fourier modes and thereby evaluate the Kelvin wave spectrum.
Since Kelvin waves can be quantized \cite{eps91}, we refer to the elementary excitations on a vortex as kelvons. The kelvon occupation number spectrum is then defined as 
\begin{equation}
n_k = \alpha_k^2 + \alpha_{-k}^2, \quad k > 0.
\end{equation}

\begin{figure}[tb] 
\begin{center} 
\includegraphics[width=\linewidth]{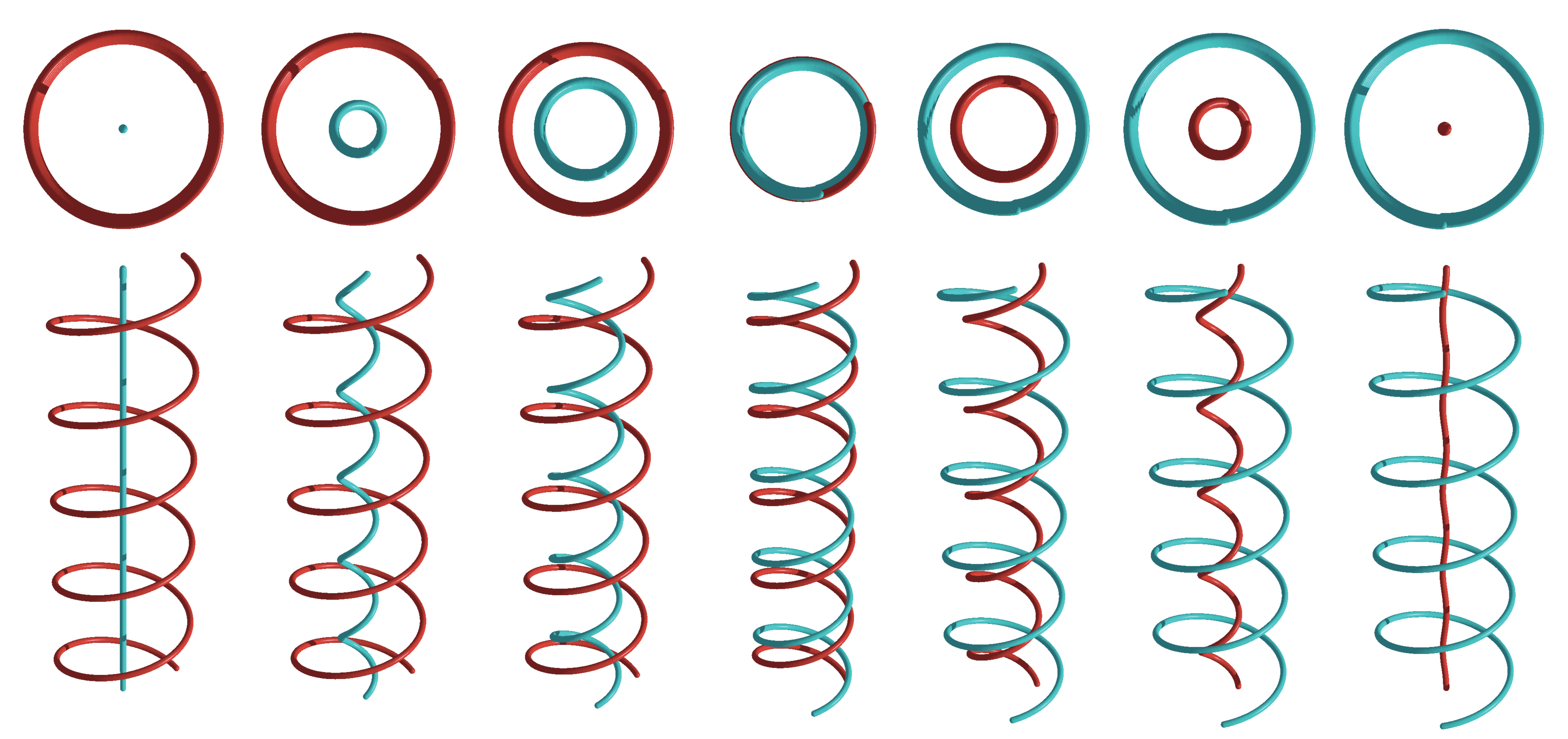}
\end{center}
\caption{Recurrence of two coaxial vortices. The initial configuration is a helical vortex with mode 5 and amplitude $A=0.015$ mm wound around a straight vortex. The length of the $z$ period is $L_z=1$ mm. Vortex configurations are shown at times
 0, 0.0035, 0.007, 0.0105, 0.014, 0.0175,  and 0.021 s. 
 Although these parameters correspond to the small-amplitude limit, the amplitudes of the Kelvin waves look large because we have used a 1:10 scaling for the transverse and axial directions of the coaxial vortices to clearly depict the helicoidal waves.
}
\label{fig:intro}
\end{figure}

The vortex configurations considered will typically have a single Kelvin wave mode. The phase of the vortex can then be obtained by simply tracking the location of point $z=0$ in the computational domain that spans the interval $0 \leq z \leq L_z$. The phase of the Kelvin wave is defined as the
azimuthal angle of that point in the $xy$~plane. On the other hand, the amplitude of the wave is determined by fitting a circle to the projection of the vortex on the $xy$~plane, the radius of that circle being the amplitude $A$. One advantage of this method is that it is also applicable to localized wave packets and localized nonlinear excitations such as Hasimoto solitons \cite{Hasimoto1972}. We have tested that our method for determining the phase and amplitude is consistent with finding them by using a fast Fourier transform.

For a helical vortex, the dimensionless product $Ak$ is an important parameter. It appears in the dispersion relation of a Kelvin wave of arbitrary amplitude (derived using the local induction approximation) \cite{son12, hie14}
\begin{equation}
\omega = \frac{\kappa}{4 \pi} \Lambda \frac{k^2}{\sqrt{1 + (Ak)^2}},
\end{equation}
where $\Lambda$ is the logarithmic prefactor of the first term containing $\hat{\bm s}' \times {\bm s}''$ in Eq. \eqref{e.bs}. The inclusion of mutual friction or counterflow would change the dispersion relation \cite{hie14}. The motion of a helical vortex is purely rotational when $Ak \rightarrow 0$. If $Ak \rightarrow \infty$, then the motion is pure translation along the vortex axis. Since the behaviour depends on the ratio of the amplitude of the wave measured with respect to the wavelength, we will use the condition $Ak = 1$ to distinguish between two different regimes.
For $Ak < 1$, we have small-amplitude Kelvin waves, whereas $Ak > 1$ corresponds to large-amplitude Kelvin waves.

\section{Numerical Results for Small amplitude Kelvin waves \label{Sec_LAKW}}

Let us consider two coaxial vortices, one straight and one with a Kelvin wave. Subsequent integration of Eq.\ \eqref{e.bs} with such an initial condition reveals that a Kelvin wave with the same wave number as that initially present on the perturbed vortex will grow on the initially straight vortex. This is compensated by a decrease in the amplitude of the initially perturbed vortex until it becomes straight. The cycle then continues, resulting in a recurrence, as can be seen in Fig. \ref{fig:intro}.  For the dynamics, see Ref. \onlinecite{supplement}.

\begin{figure}[tb] 
\begin{center} 
\includegraphics[width=0.97\linewidth]{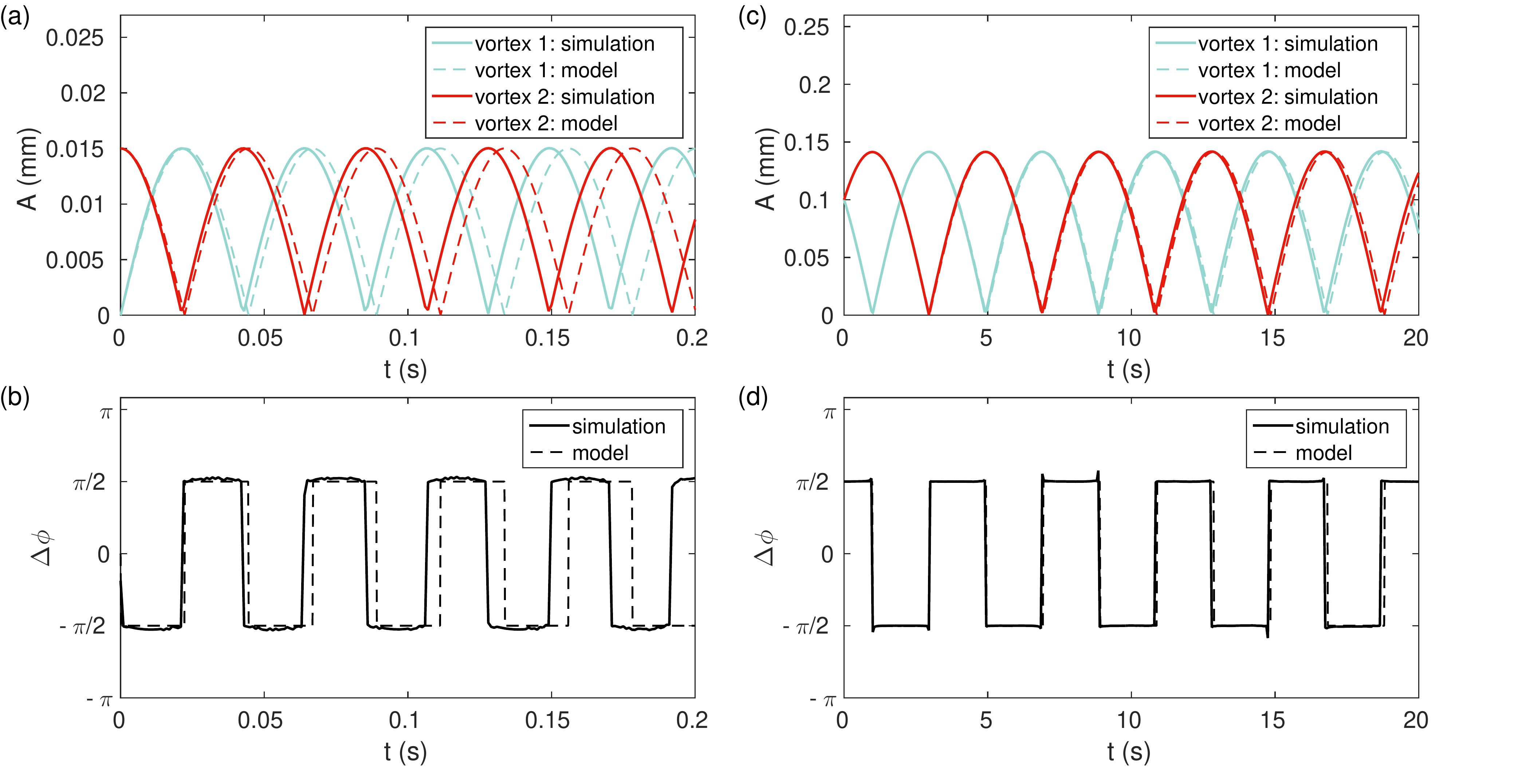} 
\end{center}
\caption{(a) and (c) Vortex amplitudes and (b) and(d) the phase difference as a function of time.  The different colors in (a) and (c) correspond to the two different vortices. Dashed lines show the solutions of the analytical model described in Sec. \ref{sec:model}. The jump in the phase occurs when one of the vortices straightens.
For (a) and (b), the initial configuration corresponds to one straight vortex, i.e. $A_1(0)=0$, and another vortex with $A_2(0)=0.015$ mm, $m=5$, and $L_z=1$ mm.
The initial configuration for the figures (c) and (d) is two vortices with equal Kelvin wave amplitudes and mode numbers ($A_1(0)=A_2(0) = 0.1$ mm, $m=1$, and $L_z=20$ mm), but with an initial phase difference $\Delta \phi = \pi/2$. In this case, $A(0)k \approx 0.03$ and so the requirement $Ak \ll 1$ is better satisfied and the difference in the time scales of the simulation and model is very small. On the other hand, the amplitude is much larger than for (a) and (b), which results in a longer recurrence time.
}
\label{fig:coaxial}
\end{figure}

When the vortices are viewed along their common axis, they appear as vortex rings passing through one another. We therefore refer to this behaviour as  leapfrogging Kelvin waves. This recurrent motion is easy to see if we consider the amplitudes of the vortices as functions of time (see Fig. \ref{fig:coaxial}). We note that the exchange of energy that is associated with the hopping of the Kelvin wave from one vortex onto the other and vice versa arises from the intrinsic nonlocal dynamics of the Biot-Savart law. The phenomena we observe cannot be described using the so-called local induction approximation, which retains only the first term in Eq.\ \eqref{e.bs}.

During the evolution, the momentum (or, more precisely, the impulse) of the vortices defined as
\begin{eqnarray}
{\bf I} = \frac{\rho_s \kappa}{2} \int {\bm s} \times  {\rm d}{\bm s}.
\end{eqnarray}
is conserved. Here $\rho_s$ is the superfluid density and the integral is performed over the interval $0 \le z \le L_z$. 
The $z$~component of the momentum (i.e., along the vortex) is proportional to the product of the area of the projection of the vortex line onto the $xy$~plane times the mode number, i.e., the number of times the vortex winds around the axis within one period. From this we can determine the instants when the amplitudes of the Kelvin waves of each vortex are equal. This should occur when $A=A(0)/\sqrt{2}$.
This is confirmed in the results presented in Fig.\ \ref{fig:coaxial}.

\begin{figure}[tb] 
\begin{center} 
\includegraphics[width=0.8\linewidth]{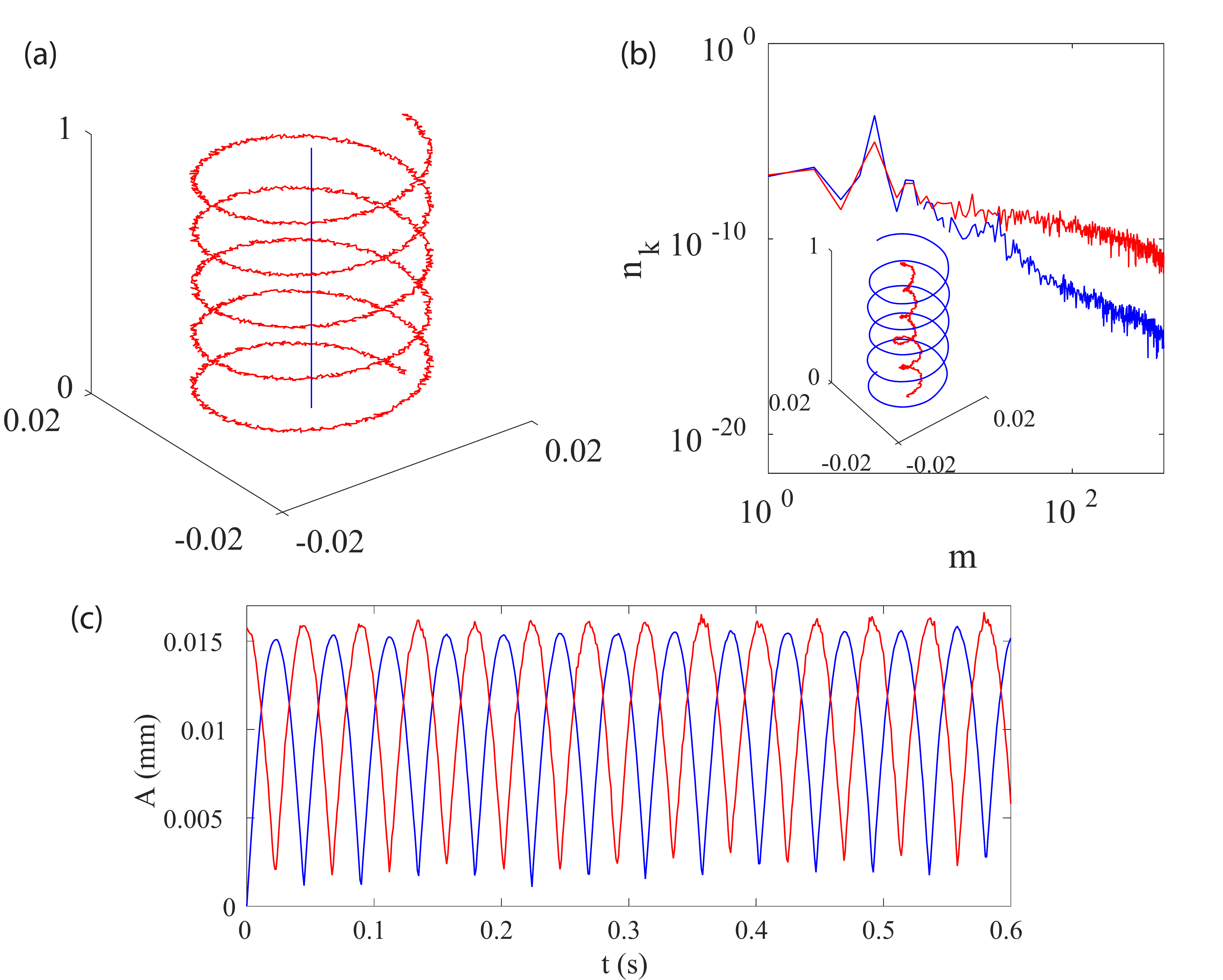}
\end{center}
\caption{Leapfrogging of Kelvin waves (for $m=5$) with random perturbations added to the initial configuration of the helical vortex.
(a) Initial configuration. 
(b) Configuration (\emph{inset}) and the kelvon occupation spectrum at $t = 0.6$ s. After more than ten recurrence periods, the initial Kelvin mode still clearly dominates. 
(c) Amplitudes of the waves as a function of time. It is apparent that the recurrence is practically unaffected by the noise.
}
\label{fig:noise}
\end{figure}

We have found that the leapfrogging motion of Kelvin waves is a robust effect, which occurs even if we add some random white noise to the initial configuration. Figure~\ref{fig:noise} shows the evolution of a Kelvin wave on a vortex to which we added some noise with magnitude equal to 2 \% of the original amplitude. It is clear that even after more than ten recurrence periods we can still see the initial Kelvin mode, confirming the persistence of the leapfrogging behaviour for initially perturbed Kelvin waves. We have also tested the stability to sideband modes ($m-1$ and $m+1$) with small amplitudes and found that such a perturbation did not destroy the recurrence of the leapfrogging.

\begin{figure}[tb] 
\begin{center} 
\includegraphics[width=0.7\linewidth]{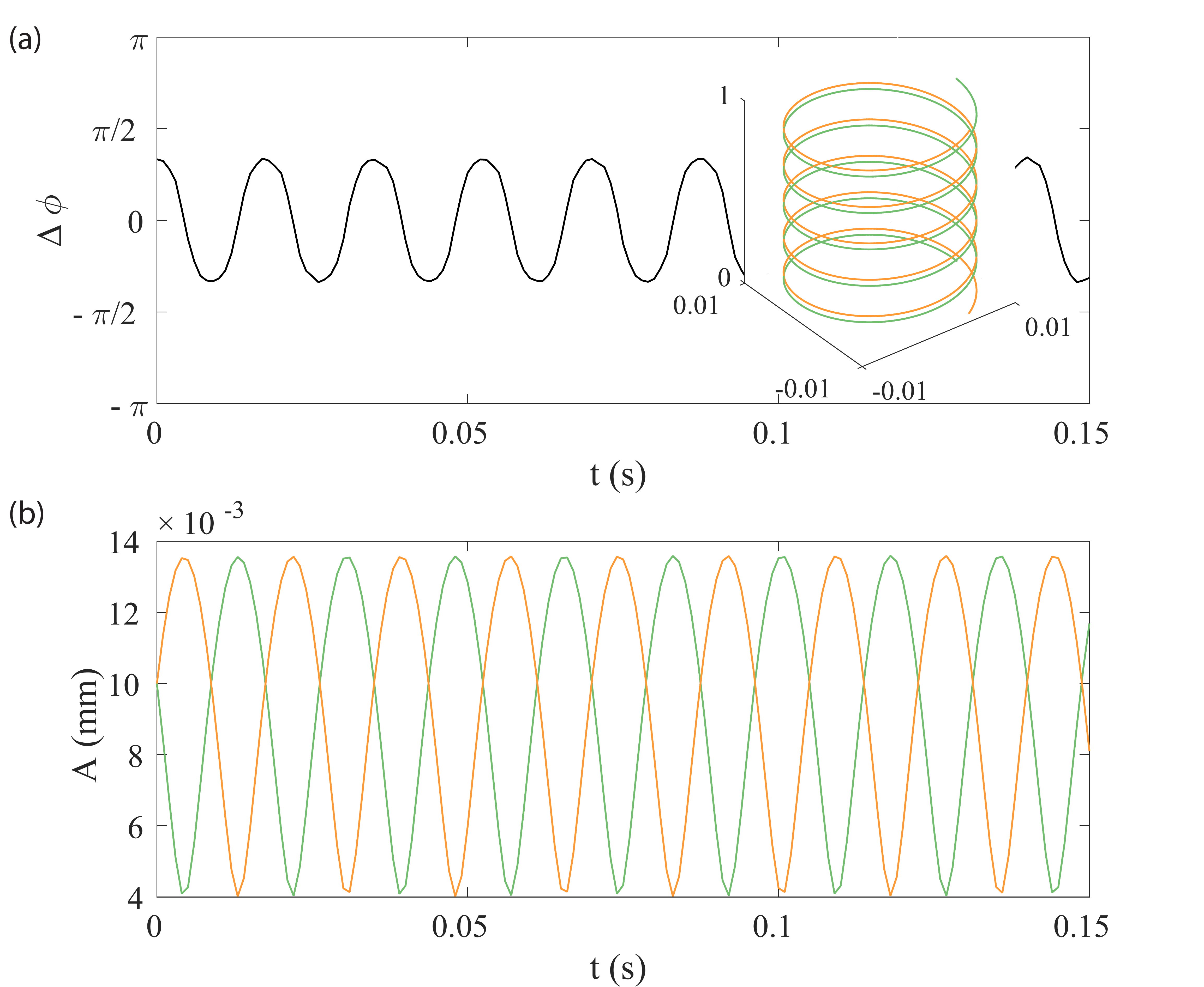}
\end{center}
\caption{(a) Time development of the phase difference and (b) Kelvin wave amplitudes. The initial phase difference is $\Delta \phi = \pi /3$, and the vortices have initially equal amplitudes of $A_1(0)=A_2(0)=0.01$ mm. The initial configuration with $m=5$, and $L_z =1$ mm is shown in the inset.
In this case the vortices do not straighten completely. 
}
\label{fig:diffPhase}
\end{figure}

\begin{figure}[hbt] 
\begin{center} 
\includegraphics[width=0.7\linewidth]{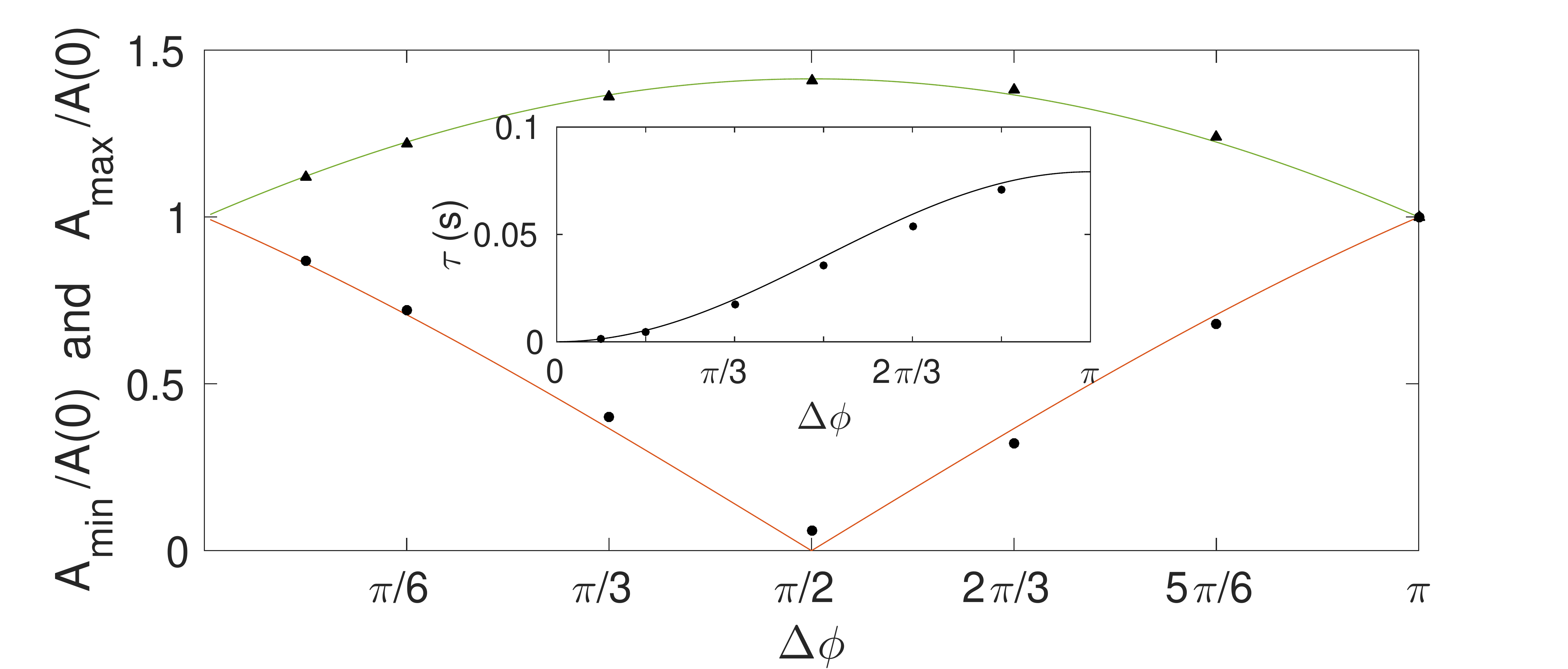} 
\end{center}
\caption{
Minimum (the red line is from the model in Sec.~\ref{sec:model} and the circles are from Biot-Savart simulations) and maximum (the green line is from the mode and the triangles are from the simulations) amplitudes for vortices, when we begin with equal amplitudes but a given phase difference ($A_1(0) = A_2(0) = 0.01$ mm, $m=5$, and $L_z = 1$ mm). 
The inset shows recurrence times as a function of the phase difference (the solid line is from the model and circles are from simulations). For $\Delta \phi = \pi$ no recurrence occurs. That is the case of a double helix, which is a steady configuration.
}
\label{fig:phase}
\end{figure}

We note that for the leapfrogging to take place, it is not necessary for one of the vortices to be initially straight. Leapfrogging occurs whenever the coaxial vortices have the same Kelvin wave mode.
If the vortices initially have equal amplitudes (an example is shown in Fig. \ref{fig:diffPhase}), then the minimum and maximum amplitudes $A_{\rm min}$ and $A_{\rm max}$ will depend on the initial phase difference (see Fig. \ref{fig:phase}). We observe that the amplitude tends to zero, i.e., the vortex straightens, only when the phase difference tends to $\pi/2$ (as in Fig. \ref{fig:coaxial}). The minimum and maximum amplitudes are symmetric with respect to $\pi /2$, while the recurrence time $\tau$ increases as the phase difference is increased. 
If the phase difference is $0< \Delta \phi \leq \pi/2$ when the vortices have equal amplitudes, then the phase difference oscillates around 0 (as can be seen from Figs. \ref{fig:coaxial} and \ref{fig:diffPhase}). In contrast, when the amplitudes are equal and $ \pi/2< \Delta \phi < \pi$, the phase difference oscillates around $\pi$.

It is remarkable that throughout the leapfrogging process, the two helical vortices do not cross and reconnect. If the phase difference is small when the vortices have equal amplitudes, they come really close to each other ($\Delta \phi$ cannot be exactly zero, since then the vortices would coincide). Even then the reconnection does not happen easily. This is because the vortices are locally parallel and any reconnection would increase the vortex length and is thus not energetically favorable. 

\begin{figure}[b] 
\begin{center} 
\includegraphics[width=0.7\linewidth]{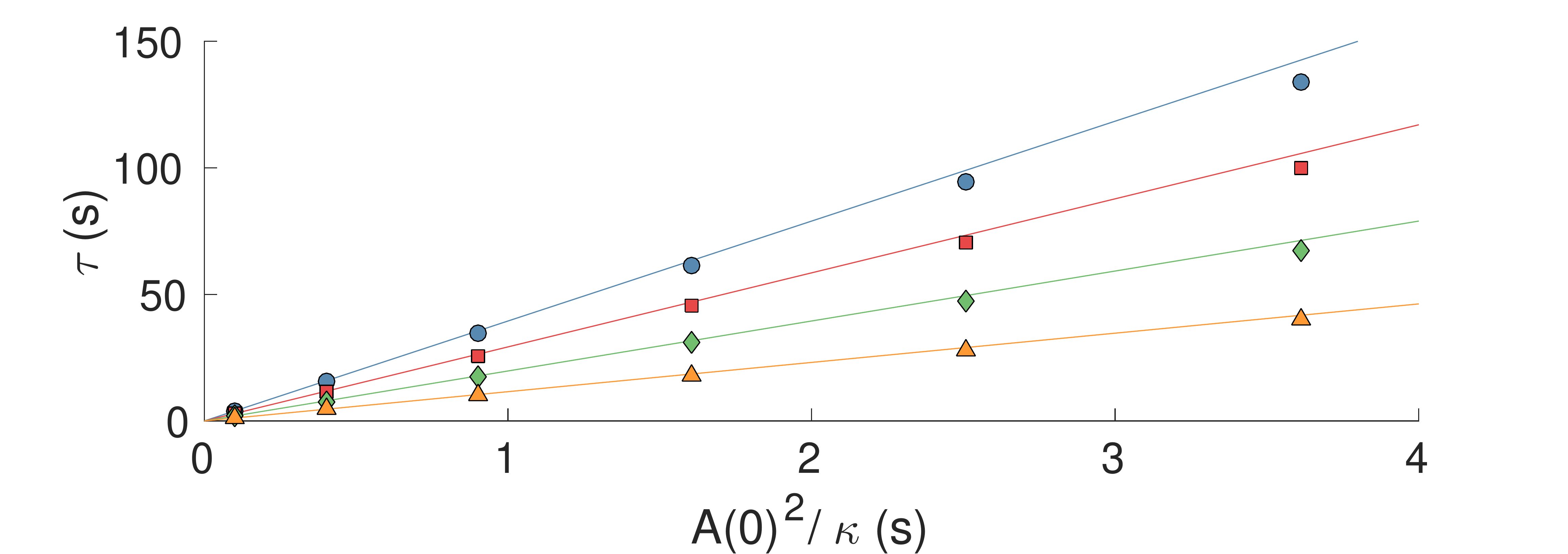} 
\end{center}
\caption{Recurrence time $\tau$ as a function of $A(0)^2 / \kappa$. The initial configuration is two vortices with equal amplitude Kelvin waves ($m=1$ and $L_z=20$ mm). The different lines are for different phase differences (blue circles, $\pi /2$; red squares, $5\pi/12$; green diamonds, $\pi/3$; and yellow triangles, $\pi/4$). Solid lines are from the model and points are from simulations. Amplitudes are between 0.1 and 0.6 mm (this corresponds to $A(0)k = 0.031, \ldots, 0.19$, so this is clearly in the small-amplitude regime). It is clear that $\tau \propto A(0)^2/\kappa$, but the proportionality coefficient depends on the phase difference.
}
\label{fig:time}
\end{figure}

\begin{figure}[hbt]
\begin{center} 
\includegraphics[width=0.85\linewidth]{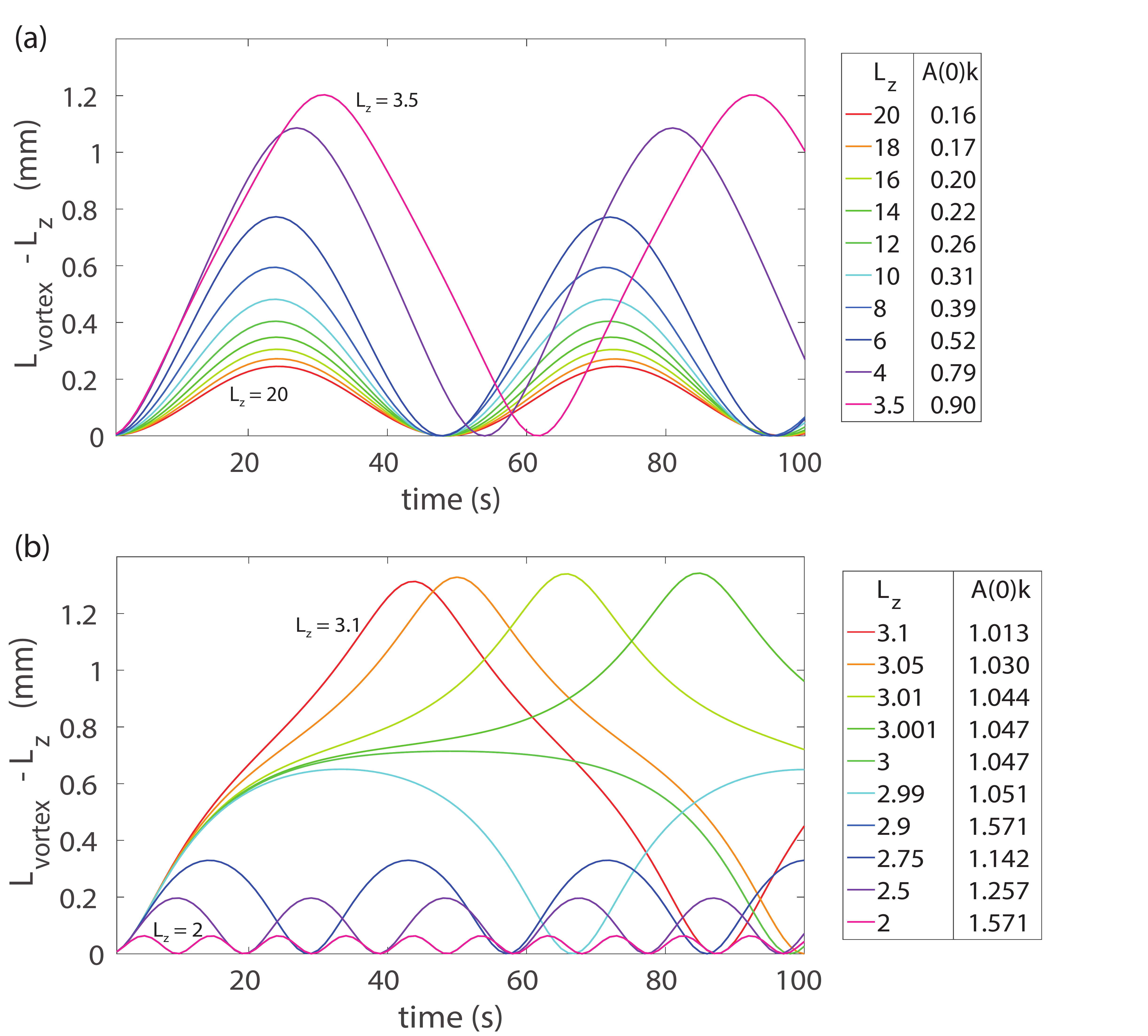} 
\end{center}
\caption{Extra length due to the Kelvin wave as a function of time. In all simulations there is one initially straight vortex and one with a Kelvin wave ($m=1$ and $A=0.5$ mm). The length of the $z$ period is varied between different runs which corresponds to a variation in the wavelength of the Kelvin wave. Only the length of the initially straight vortex is plotted. 
(a) For all of the small-amplitude cases ($A(0)k < 1$ or $L_z > 2\pi A(0) = \pi$ mm) the recurrence time is almost the same. Deviations are most visible when $A(0)k \sim 1$. 
(b) For $A(0)k > 1$ (or $L_z < 2\pi A(0) = \pi$ mm) the recurrence times do not follow the same pattern as for the long wavelengths. For $L_z \leq 3$ mm the amplitudes of the two vortices are never equal. 
}
\label{fig:wavelength}
\end{figure}

If we ignore the wavelength, then $A(0)$ and $\kappa$ are the only relevant dimensional parameters in the system and they can be combined to identify a relevant time scale. Indeed, the recurrence time $\tau$ seems to be proportional to $A(0)^2/\kappa$ for a given phase difference (see Fig.\ \ref{fig:time}). In Fig.\ \ref{fig:time} we included only numerical simulations with very small amplitudes for practical computational reasons. Figure~\ref{fig:wavelength}(a) also shows that for vortices with small-amplitude Kelvin waves, the recurrence time depends only very weakly on the wavelength of the Kelvin wave. However, in Fig.\ \ref{fig:wavelength} we can see that when $Ak$ is close to unity, the recurrence time is no longer proportional to $A(0)^2/\kappa$.

\section{Model of Leapfrogging for Nearly parallel vortices} \label{sec:model}
\vspace{1ex}

We have seen that helical vortices can be expressed in the form of Eq. \eqref{eq.w}. If the amplitude is small, the helices are nearly parallel to the $z$ axis. In this case, it is possible to describe the motion of the vortices using simplified equations that account for the interaction of nearly parallel vortex filaments as given in Refs. \onlinecite{kle95,zak99}. This model essentially combines a local induction approximation (LIA)  for each vortex with a nonlocal term describing the interaction with the other vortices 
in a layered fashion.

To provide a systematic but more straightforward derivation of this model, we will begin by considering the Hamiltonian formulation of the Biot-Savart law as given in Ref. \onlinecite{koz09}. In particular, we recall that the Biot-Savart law conserves energy as given by
\begin{eqnarray}
E = \sum_{j,k=1}^{N_v,N_v} \frac{\kappa_j \kappa_k}{4\pi}  \iint \frac{{\rm d}{\bm s}_j \cdot {\rm d}\tilde{{\bm s}}_{k}}{|{\bm s}_j-\tilde{{\bm s}}_{k}|},
\end{eqnarray}
where the double summation runs over all the $N_v$ vortices.
From this definition for the energy, it follows that for a filament that can be parameterised in terms of the $z$ coordinate, the Hamiltonian can be expressed as
\begin{eqnarray}
H = \sum_{j,k=1}^{N_v,N_v} \frac{\kappa_j \kappa_k}{4\pi} \iint \frac{[1+{\text Re}\{ w_j'^*(z_1)w_k'(z_2) \}]{\rm d}z_1 {\rm d}z_2}{\sqrt{(z_1-z_2)^2+|w_j(z_1)-w_k(z_2)|^2}}.
\end{eqnarray}
Here the prime in $w'(z)$ denotes differentiation with respect to the argument $z$.
Since the integrand in the above expression for $H$ is singular as $z_1 \rightarrow z_2$ for $j=k$, we regularise the integrand by introducing cutoff scales
$l_{\pm}$ as described in Sec.~\ref{Sec_Methods} such that $a_0 \ll l_{\pm} \ll 1/c$. This choice of the cutoff scale, which has also been described in Ref. \onlinecite{sch85}, leads to
\begin{eqnarray}
H \approx & \sum_{j=1}^{N_v} \frac{\kappa_j^2}{2\pi} \ln\left( \frac{2\sqrt{l_{+}l_{-}}}{e^{1/2}a_0} \right) \int {\rm d}z \sqrt{1+|w_j'(z)|^2} \\
& + \sum_{j,k=1}^{N_v,N_v} \frac{\kappa_j \kappa_k}{4\pi} \iint' \frac{[1+{\text Re}\{ w_j'^*(z_1)w_k'(z_2) \}]{\rm d}z_1 {\rm d}z_2}{\sqrt{(z_1-z_2)^2+|w_j(z_1)-w_k(z_2)|^2}}, \nonumber
\end{eqnarray}
where, as before, the prime on the integrals implies that the integrals exclude the interval ${|z_1-z_2|>\sqrt{l_+ l_-}}$ when $j=k$ (i.e., the contribution to the self-energy of a vortex). We will restrict our analysis to two coaxial corotating vortices with circulation $\kappa_j = \kappa_k = \kappa$. Moreover, if we assume small-amplitude Kelvin wave perturbations on the vortices with amplitude $A$ and wave number $k$, then $w'(z) \ll 1$, which is equivalent to our earlier assumption that $Ak = \epsilon \ll 1$. Under this assumption and considering the case where $j=k$, we find that the integrand in the nonlocal contribution to the Hamiltonian can be approximated to leading order by
\begin{eqnarray}
 \iint' \frac{[1+{\text Re} \{ w_j'^*(z_1)w_j'(z_2) \}]{\rm d}z_1 {\rm d}z_2}{\sqrt{(z_1-z_2)^2+|w_j(z_1)-w_j(z_2)|^2}} 
 \approx \iint' \frac{{\rm d}z_1 {\rm d}z_2}{\sqrt{(z_1-z_2)^2}}.
\end{eqnarray} 
In arriving at the above approximate form on the right-hand side, we have made use of the fact that $|z_1-z_2|\ge \sqrt{l_{+} l_{-}}$ since the prime on the double integral assumes that a small interval is excluded for $j=k$. Since $Ak \ll 1$ and $l_{\pm} \ll 1/c$, we can satisfy both of these conditions with the further assumption that $A<l_{\pm}$. It follows that $|w_j(z_1)-w_j(z_2)|/|z_1-z_2|<1$ and we can neglect this latter term in comparison to the $(z_1-z_2)$ term from which the above expression follows.

Since the leading-order expression does not depend on the function $w(z)$ or its derivatives, this term plays no role in the motion of the vortices. On the other hand, for $j \ne k$, we transform the integration variables to
\begin{eqnarray}
z^+ = z_1+ z_2, \quad z^- = z_1 - z_2.
\end{eqnarray}
Taylor expanding the complex functions $w$ about the point $z^+$, we have
\begin{eqnarray}
w_j(z^++z^-) = w_j(z^+) + w_j'(z^+) z^-  + w_j''(z^+) \frac{(z^-)^2}{2} +  \cdots \text{,} \nonumber \\
w_k(z^+-z^-) = w_k(z^+) - w_k'(z^+) z^-  + w_k''(z^+) \frac{(z^-)^2}{2} +  \cdots \text{.} \nonumber
\end{eqnarray}
After neglecting the $w'(z)$ terms in the numerator due to the assumption of small-amplitude Kelvin waves, this gives
\begin{equation}
 \iint \frac{[1+{\text Re} \{ w_j'^*(z_1)w_k'(z_2) \}]{\rm d}z_1 {\rm d}z_2}{\sqrt{(z_1-z_2)^2+|w_j(z_1)-w_k(z_2)|^2}} 
\approx \iint \frac{2 {\rm d}z^- {\rm d}z^+}{\sqrt{(z^-)^2+|w_j(z^++z^-)-w_k(z^+-z^-) |^2}}.
\end{equation}
In order to identify the leading-order contributions, we focus on the integral over $z^-$ and perform a scaling analysis by splitting the integral into an inner integral extending over the interval $\mathcal{C}_i = \{z^-: |z^-| \le \sqrt{A/k} \}$ and an outer interval for $\mathcal{C}_o = \{z^-:|z^-| >\sqrt{A/k}\}$. The intermediate length scale $\sqrt{A/k}$ is chosen such that it is large relative to the amplitude $A$ that is used as the length scale for the inner integral, but is small in comparison to the wavelength $2\pi/k$ that is used to define the length scale for the outer integral. Hence, introducing the rescaled variables $\tilde{z} = z/A$, and $\tilde{w}^{(p)} = w^{(p)}/(Ak^p)$ for the inner integral and $Z = zk$, and $W = w^{(p)}/(Ak^p)$ for the outer integral, where $w^{(p)}$, $p=0,1,2,\cdots$, denotes the $p$th derivative of $w$, we obtain
\begin{eqnarray}
&\phantom{=}& \left( \int_{\mathcal{C}_o}  + \int_{\mathcal{C}_i} \right) \frac{2 {\rm d}z^-}{\sqrt{(z^-)^2+|w_j(z^++z^-)-w_k(z^+-z^-) |^2}} \nonumber \\
&=&  \left( \int_{\epsilon^{1/2}}^{\infty} + \int_{-\infty}^{-\epsilon^{1/2}} \right) \frac{2 {\rm d}Z^-}{\sqrt{(Z^-)^2+ \epsilon^2 |W_j-W_k + \cdots|^2}} \nonumber \\
&+& \int_{-\epsilon^{-1/2}}^{\epsilon^{-1/2}} \frac{2 {\rm d}\tilde{z}^-}{\sqrt{(\tilde{z}^-)^2+|\tilde{w}_j-\tilde{w}_k + \epsilon( \tilde{w}_j'+\tilde{w}_k') \tilde{z}^- + \cdots|^2}} \nonumber \\
&\approx&  \int_{-\epsilon^{-1/2}}^{\epsilon^{-1/2}} \frac{2 {\rm d}\tilde{z}^-}{\sqrt{(\tilde{z}^-)^2+|\tilde{w}_j-\tilde{w}_k |^2}} + \text{constant} \nonumber \\
&\approx&  -2\ln |w_j(z^+)-w_k(z^+)| + \text{constant}.
\end{eqnarray}
Note that the limits  tend to infinity for the integrals over the inner coordinates as $\epsilon$ tends to zero, while the separation scale tends to zero for the integral over the outer coordinates.
It follows that after expanding the square root term in the local term (LIA contribution) 
to first order, neglecting the constant terms, and redefining $z^+ \rightarrow z$, the total Hamiltonian to low-amplitude Kelvin waves can then be expressed as
\begin{eqnarray}
H \approx \int \left( \sum_{j=1}^{N_v} \frac{\kappa^2 \Lambda}{4\pi} |w_j'(z)|^2
- \sum_{\substack{j,k=1 \\ j\neq k}}^{N_v,N_v} \frac{\kappa^2}{2\pi} \ln{|w_j(z)-w_k(z)|} \right) {\rm d}z. \nonumber
\end{eqnarray}
We note that since the LIA term contains an independent parameter given by $\ln(\sqrt{l_+ l_-}/a_0)$, the two terms in the above expression for the Hamiltonian are equally important in the distinguished limit corresponding to $\epsilon^2 \Lambda \sim 1$.

We have approximated the logarithmic factor to be constant and for helium-4 its value in typical experiments is $\Lambda \simeq 12$.  
We remark that since the LIA ignores all nonlocal interactions it alone cannot explain the interaction between the two vortices. 
 In this model, the nonlocal interactions are approximated to be similar to straight line vortices interacting in a layered fashion such that points of the vortex filaments lying at the same value of $z$ interact as though they were point vortices lying in a plane. We note that the model we have derived was also obtained by Klein \emph{et al.} in Ref. \onlinecite{kle95}, although our derivation and application differ from theirs. An intuitive explanation for the above form of the Hamiltonian  that we have derived is that since the vortices are almost straight,
their respective self-induced velocities must be very small. On the other hand, since each vortex is also nearly parallel to the $z$~axis, the velocity it induces on the other is inversely proportional to their separation. 

We can now recover the equations of motion for each vortex filament by using
\begin{equation}
i \frac{\partial w_j}{\partial t} = \frac{1}{\kappa_j} \frac{\delta H[w]}{\delta w_j^*} \text{.}
\end{equation}
Hence, evaluating the equations of motion from the above simplified form of the Hamiltonian, we recover
\begin{equation} \label{e.klein}
\frac{\partial w_j}{\partial t} = i\frac{\kappa \Lambda}{4 \pi} \frac{\partial^2 w_j}{\partial z^2}  + i\frac{\kappa}{2 \pi} \frac{w_j-w_k}{|w_j-w_k|^2} \text{,} 
\end{equation}
where $j=1$ when $k=2$ or vice versa. If we introduce the coordinates $u=w_1 - w_2$ and $v = w_1 + w_2$, Eq.\ \eqref{e.klein} transforms to
\begin{eqnarray}
\frac{\partial u}{\partial t} &=& i\frac{\kappa \Lambda}{4 \pi}\frac{\partial^2 u}{\partial z^2} + i\frac{\kappa}{2 \pi}\frac{2u}{|u|^2}, \\
\frac{\partial v}{\partial t} &=& i\frac{\kappa \Lambda}{4 \pi}\frac{\partial^2 v}{\partial z^2} \text{.}
\end{eqnarray}
These equations admit plane-wave solutions for both $u$ and $v$, which are given by
\begin{eqnarray}
u &=& Be^{i(kz-\omega_ut+\theta_u)}, \quad v = Ce^{i(kz-\omega_vt+\theta_v)}, \\
\omega_u &=& \frac{\kappa \Lambda}{4 \pi}k^2 - \frac{\kappa}{2 \pi}\frac{2}{B^2} \quad \text{and} \quad \omega_v = \frac{\kappa \Lambda}{4 \pi}k^2 \text{,}
\end{eqnarray}
where $B$ and $C$ are real constants with units of length. Without loss of generality, we can restrict $B$ and $C$ to be positive since the negative values can be encoded within the phases $\theta_u$ and $\theta_v$. Alternatively, we can set $\theta_u = \theta_v =0$ because these constants will simply shift the origin of the graphs for the amplitudes and phases. In principle, the wave number $k$ could be different for $u$ and $v$, but the case with the same $k$ is relevant for the leapfrogging Kelvin waves that are the focus of this work. The corresponding expressions for the complex coordinates of the two vortices are given by
\begin{eqnarray}
 w_1 &=& \frac{C}{2} e^{i(kz-\omega_v t)} + \frac{B}{2} e^{i(kz-\omega_u t)} = A_1(t) e^{ikz} e^{i\phi_1(t)} \text{,} \\
 w_2 &=& \frac{C}{2} e^{i(kz-\omega_v t)} - \frac{B}{2} e^{i(kz-\omega_u t)} = A_2(t) e^{ikz} e^{i\phi_2(t)} \text{,}
\end{eqnarray}
where the amplitudes are
\begin{equation}
A_{1,2}(t) = \frac{1}{2} \sqrt{C^2 +B^2 \pm 2BC \cos\bm( t(\omega_u - \omega_v)\bm)}, \quad 
\end{equation}
and the phases are given by
\begin{equation}
\tan \phi_{1,2}(t) = -  \frac{C\sin \omega_v t \pm B \sin \omega _u t}{C\cos \omega_v t \pm B \cos \omega _u t} \text{.}
\end{equation}
In the above equations, the positive and negative signs correspond to $w_1$ and $w_2$, respectively.

Solutions of this type represent two leapfrogging coaxial helical vortices with initial amplitudes $A_1(0) = \frac{1}{2}|C+B|$ and $A_2(0)=\frac{1}{2} |C-B|$ and initial phase difference $0$ or $\pi$. Due to our choice of the initial phase difference (i.e., setting $\theta_u = \theta_v =0$), the amplitudes $A_1(0)$ and $A_2(0)$ also correspond to the maximum and minimum amplitudes. The recurrence time is then given by
\begin{equation}
\tau = \frac{2 \pi}{\omega_v - \omega_u} = \frac{2 \pi^2 B^2}{\kappa} \text{.}
\end{equation}
In Sec.~\ref{Sec_LAKW} we arrived at the proportionality $\tau \propto A(0)^2/\kappa$ from dimensional reasoning. In Fig. \ref{fig:time} we show $\tau$ as a function of $A(0)^2/ \kappa$, where $A(0)$ corresponds to the value of the amplitude at the instant when the respective amplitudes of the Kelvin waves of both vortices become equal. 
This is true because $\tau \propto B^2$, where $B=\sqrt{4A(0)^2 - C^2}$.

This model gives very good agreement with the results of our numerical simulations of the vortex filament model. We note that since $B$ and $C$ determine the phase difference, $\tau$ also depends on $\Delta \phi$, as observed in our simulations. In Fig. \ref{fig:coaxial}, the results of both the numerical simulation and the solution of the model are shown.   For the case with $A(0)k \approx 0.47$ presented in Figs.\ \ref{fig:coaxial}(a) and \ref{fig:coaxial}(b), we note that the time scale of the recurrence predicted by the model differs slightly from the numerical solution. We attribute this to the value of $Ak$ used in this case,
which is not sufficiently small to agree with the assumptions made in arriving at Eq.\ \eqref{e.klein}. The second simulation presented in Figs.\  \ref{fig:coaxial}(c) and \ref{fig:coaxial}(d) corresponding to $A(0)k \approx 0.03$ satisfies the condition $Ak \ll 1$ better and consequently we observe better agreement with the prediction of our model. Therefore, in the small-amplitude limit, the accuracy of the approximations made improves.

\section{Large amplitude Kelvin waves}

\begin{figure}[tb]
\begin{center} 
\includegraphics[width=0.7\linewidth]{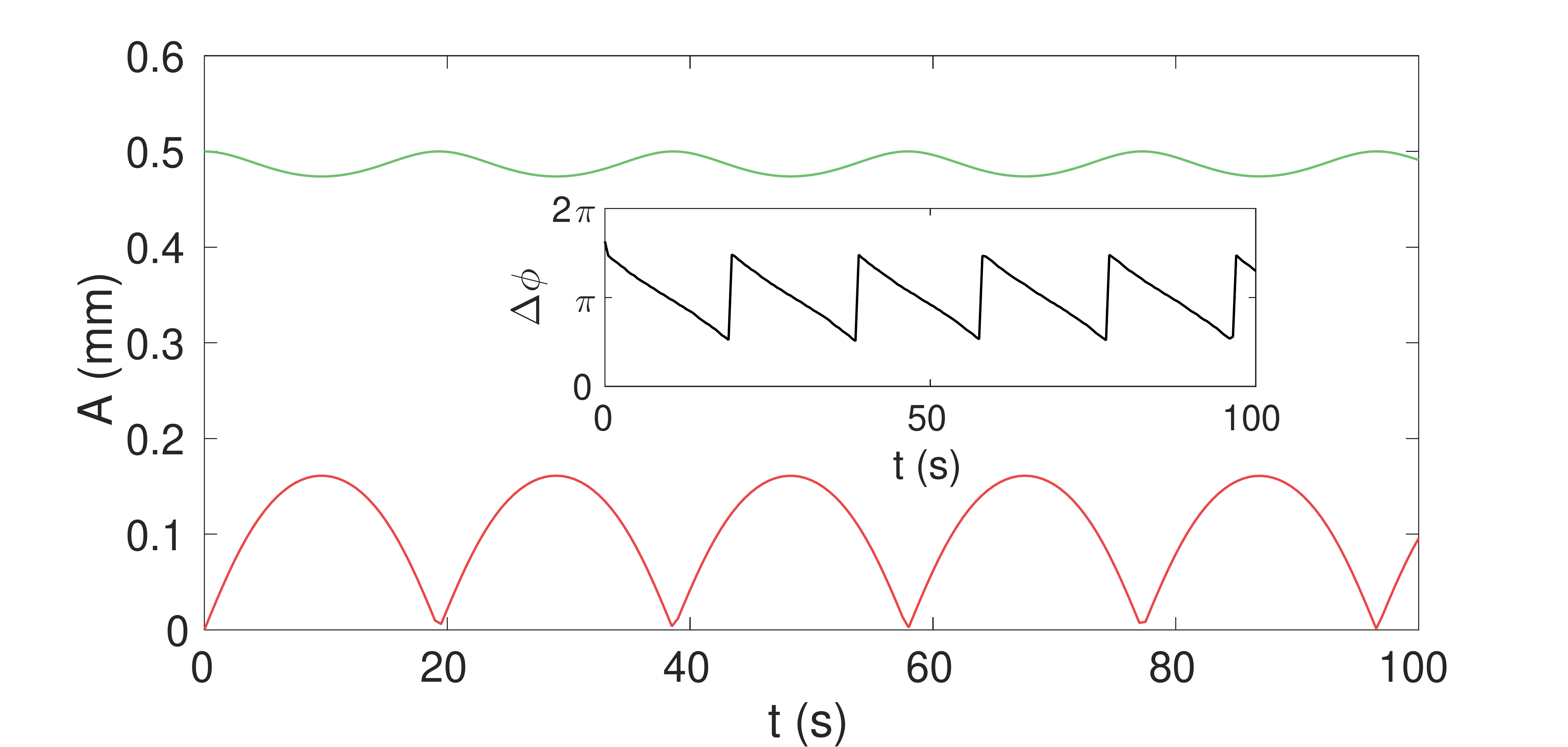} 
\end{center}
\caption{Amplitudes and phase difference (\emph{inset}) as a function of time. Initially, one of the vortices is straight and the other vortex has a Kelvin wave ($m = 1$, $A = 0.5$ mm, and $L_z = 2.5$ mm). In the large-amplitude regime, the amplitudes oscillate within different ranges. 
}
\label{fig:large}
\end{figure}

In the preceding sections we have focused on small-amplitude Kelvin waves and derived a model to explain the observed recurrence phenomena in this limit. We will now consider what happens if we relax the assumption of small-amplitude Kelvin waves. The model we have derived is no longer applicable in this parameter regime and we will proceed by relying on numerical simulations alone. 
To identify qualitative differences in the dynamics of the vortices, we have tracked how the length of one of the vortices in our simulations changes with time  (see Fig.\ \ref{fig:wavelength}). This provides sufficient insight into the dynamics because, for a helix, the amplitude and length are related through the expression $L_{\rm helix} = L_z\sqrt{1+(Ak)^2}$. The amplitude of the other vortex can then be recovered by using momentum conservation. The results of our simulations, presented in Fig. \ref{fig:wavelength}, reveal that if the amplitude is increased, the frequency of the recurrence is modified, indicating that the vortex motion is being influenced by nonlinear effects. Moreover, for sufficiently large amplitudes, there is a qualitative change in the behaviour of the system, indicating a transition from one type of recurrent motion to another. In addition, if the Kelvin wave amplitude of one of the vortices is sufficiently large in comparison with the other, then the amplitudes of the waves on the vortices might never coincide with one another. 
In this case, the amplitudes of the Kelvin waves in each vortex will oscillate within different ranges (see Fig. \ref{fig:large} and Ref.~\onlinecite{supplement}). 

In order to explain the difference between the case of small amplitudes and the case of large amplitudes, it is useful to draw the following analogy.
If we consider two coaxial vortex rings, 
 it is possible that one of the rings escapes and no leapfrogging occurs. In the case of Kelvin waves, there is no direct analogy to this effect, so we must consider two rows of coaxial vortex rings instead of just two vortex rings. Now there are two possibilities. The first possibility takes place if rings of the inner row leapfrog with corresponding rings of the outer row. This scenario is analogous to leapfrogging of small-amplitude Kelvin waves. The analogy then allows us to associate the amplitude and wavelength of the Kelvin wave with the radius of the rings and the distance between consecutive rings on the same row, respectively. Moreover, the phase difference at the instant when the amplitudes of the vortices are equal is related to the distance between the rings in the two rows at the moment when they have equal radii.

For the scenario of large-amplitude Kelvin waves, the analogy with two coaxial rows of vortex rings implies that a qualitatively different behaviour is expected. In this case, if the ring in the inner row escapes the corresponding outer ring, its radius may increase some time until it approaches the next ring in the outer row. Then the inner ring shrinks and passes through the outer ring and this cycle repeats. This analogy can provide a qualitative interpretation for the change in leapfrogging of Kelvin waves between small amplitudes and large amplitudes. Furthermore, it provides an explanation for the sudden qualitative difference in the recurrence phenomena as the amplitudes are increased. However, care must be taken not to push this analogy too far as it cannot provide accurate quantitative predictions. In particular, rings having amplitudes and distances that are similar to Kelvin wave amplitudes and wavelengths will not behave in exactly the same way.

\section{Conclusions}
We have shown that coaxial vortices with the same Kelvin wave mode exchange energy in a periodic fashion. The helical vortices move through each other in a way that is similar to leapfrogging vortex rings. In the small-amplitude limit, the amplitudes of the two helical waves oscillate within the same range. In contrast, in the large-amplitude limit, the amplitudes oscillate around two different mean values, although leapfrogging continues to occur in a periodic fashion. 

Using a simplified model for nearly parallel vortices, we were able to explain the leapfrogging behaviour  observed in our Biot-Savart simulations in the limit of small-amplitude Kelvin waves. The success of the model in this regime suggests that it may be possible to generalize it to other contexts, most notably in studying how adjacent vortex filaments can influence the theoretically predicted Kelvin wave cascades in the high-wave-number regime of superfluid turbulence. For example, such a scenario would be relevant to understanding Kelvin waves on polarized vortex bundles that are believed to form in superfluid turbulence.

In the large-amplitude Kelvin wave regime, there is a qualitative change in the dynamical behaviour of the vortices. We have been able to interpret this by  drawing an analogy with the dynamics of coaxial rows of vortex rings. We end by noting that the phenomena predicted here will also contribute to our understanding of the dynamics of vortices and their interaction in classical fluids.


\begin{acknowledgments}
R.H.\ and N.H.\ acknowledge financial support from the Academy of Finland. H.S.\ acknowledges support for a Research Fellowship from the Leverhulme Trust under Grant R201540. N.H.\ thanks CSC -- IT Center for Science, Finland, for computational resources.
\end{acknowledgments}

\end{document}